\documentclass[aps,prl, showpacs, twocolumn, superscriptaddress,reprint]{revtex4-1}
\usepackage{graphicx}
\usepackage{bm}
\usepackage{gensymb}

\begin{document}



\title{Optically driven spin pumping mediating collective magnetization dynamics \\
in a spin valve structure}

\author{A. P. Danilov}
\affiliation{Experimentelle Physik 2, Technische Universit\"at Dortmund, D-44227 Dortmund, Germany}

\author{A. V. Scherbakov}
\affiliation{Experimentelle Physik 2, Technische Universit\"at Dortmund, D-44227 Dortmund, Germany}
\affiliation{Ioffe Institute, 194021 St.Petersburg, Russia}

\author{ B. A. Glavin}
\affiliation{Department of Theoretical Physics, V.E. Lashkaryov Institute of Semiconductor Physics, 03028 Kyiv, Ukraine}

\author{T. L. Linnik}
\affiliation{Department of Theoretical Physics, V.E. Lashkaryov Institute of Semiconductor Physics, 03028 Kyiv, Ukraine}

\author{A. M. Kalashnikova}
\affiliation{Ioffe Institute, 194021 St.Petersburg, Russia}

\author{L. A. Shelukhin}
\affiliation{Ioffe Institute, 194021 St.Petersburg, Russia}

\author{D. P. Pattnaik}
\affiliation{School of Physics and Astronomy, University of Nottingham, Nottingham NG7 2RD, UK}

\author{A.~W.~Rushforth}
\affiliation{School of Physics and Astronomy, University of Nottingham, Nottingham NG7 2RD, UK}

\author{C.~J.~Love}
\affiliation{Department of Physics, University of York, York YO10 5DD, UK}

\author{S. A. Cavill}
\affiliation{Department of Physics, University of York, York YO10 5DD, UK}
\affiliation{Diamond Light Source Chilton, Didcot, Oxfordshire OX11 0DE, UK}

\author{D. R. Yakovlev}
\affiliation{Experimentelle Physik 2, Technische Universit\"at Dortmund, D-44227 Dortmund, Germany}
\affiliation{Ioffe Institute, 194021 St.Petersburg, Russia}


\author{M. Bayer}
\affiliation{Experimentelle Physik 2, Technische Universit\"at Dortmund, D-44227 Dortmund, Germany}
\affiliation{Ioffe Institute, 194021 St.Petersburg, Russia}

\begin{abstract}
We demonstrate spin pumping, i.e. the generation of a pure spin current by precessing magnetization, without application of microwave radiation commonly used in spin pumping experiments. We use femtosecond laser pulses to simultaneously launch the magnetization precession in each of two ferromagnetic layers
of a Galfenol-based spin valve 
and monitor the temporal evolution of the magnetizations. The spin currents generated by the precession cause a dynamic coupling of the two layers. This coupling has dissipative character and is especially efficient when the precession frequencies in the two layers are in resonance, where coupled modes with strongly different decay rates are formed.\end{abstract}

\maketitle

The generation of a spin current (SC) by magnetization precession (MP) is known as spin pumping (SP) \cite{1}. Thereby, the precessing magnetization of a ferromagnetic (FM) film transfers angular momentum to an adjacent material, representing a pure SC that is not accompanied by the flow of charges. SCs generated by SP contain an ac-component at the precession frequency and carry also the MP phase. Conceptually, SP offers a new way of building spintronic devices by flexibly combining conducting and insulating materials \cite{2,3,4,5,6,7,8}. This has stimulated intense efforts aimed at demonstrating SCs in a robust way \cite{9}.

	Conventional SP experiments exploit a ferromagnetic resonance (FMR) where the MP is driven by a microwave field \cite{10}. The transfer of angular momentum to the adjacent material results in enhanced damping of the FMR \cite{11,12} and thus to a broadening of the corresponding resonance spectrum \cite{13,14}. In turn, the SC injected into the adjacent layer can be detected by, for example, the inverse spin Hall effect \cite{2,3,4,5,6,7,8,15,16,17,18,19,20,21,22}. In a spin valve structure consisting of two FM layers separated by a non-magnetic spacer, the SC generated by one layer drives the magnetization precession of the other layer \cite{23,24,25,26}. At resonance, when the precession frequencies of the FM layers coincide, a strongly coupled collective precessional mode forms \cite{27,28}.

	This conventional approach has a drawback, however: applying monochromatic microwave fields for driving the MP lacks the flexibility required for nanoscale applications, it strictly sets the MP and SC phase, and requires exact matching to the FMR frequency. Ultrafast optical excitation, widely used nowadays in ultrafast optomagnetism for launching MPs \cite{29}, is a promising alternative. In metallic FMs, ultrashort laser pulses trigger MP by rapidly alternating the magnetic anisotropy \cite{30}. While laser pulses have been utilized for SC generation via the transport of spin-polarized electrons from an optically-excited magnetic region \cite{31,32,33,34,35,36}, no evidence of pure SCs generated by optically launched MP has been reported.

	In this Letter, we report optically excited SP in a pseudo spin-valve (PSV) consisting of two FM layers separated by a normal metal spacer. By femtosecond laser pulses we simultaneously excite MP in the two magnetic layers. We unambiguously demonstrate that the mutual SP modifies the precession dynamics, as evidenced by strongly coupled resonant MP. In contrast to microwave driven methods
ultrafats optical excitation and 
time-resolved detection allows us to create 
a superposition of two degenerate precessional modes with split decay rates, which indicates strong dissipative coupling rarely observed experimentally.

      Figure 1a sketches the experiment. The structure under study is a PSV based on Galfenol, an alloy of iron and gallium, which possesses large saturation-magnetization and narrow FMR linewidth \cite{37}. In addition, Galfenol is a material with strong magnetoelastic coupling, which allows efficient excitation of MP by both thermal and acoustic mechanisms \cite{38}. The structure was grown epitaxially  on a (001)-GaAs substrate and contains two Fe$_{0.81}$Ga$_{0.19}$ layers: one layer of 4-nm width (Layer 1) was deposited directly on GaAs; the other 7-nm wide layer (Layer 2) was separated from the first one by a copper spacer of 5-nm thickness. The structure was covered by a 150-nm SiO$_2$ protective cap. The 5-nm Cu-thickness prevents indirect exchange interaction between the two FM layers \cite{39}. Their magnetizations, $\bm{\mbox{M}}_1$ and $\bm{\mbox{M}}_2$, can be aligned by an external magnetic field,
based on their magnetic anisotropies. Figure~1b shows the magnetization curves measured by SQUID magnetometry for three in-plane directions of the external field, $\bm{\mbox{B}}$, which are described by the azimuthal angle $\varphi_B$ 
(see insert). The easy axes of both layers are along the [100] crystal direction ($\varphi_B=0\degree$). At $B>50$~mT the structure is fully 
saturated along $\bm{\mbox{B}}$ with $\bm{\mbox{M}}_1 || \bm{\mbox{M}}_2$.

\begin{figure}
 \includegraphics[scale=0.55]{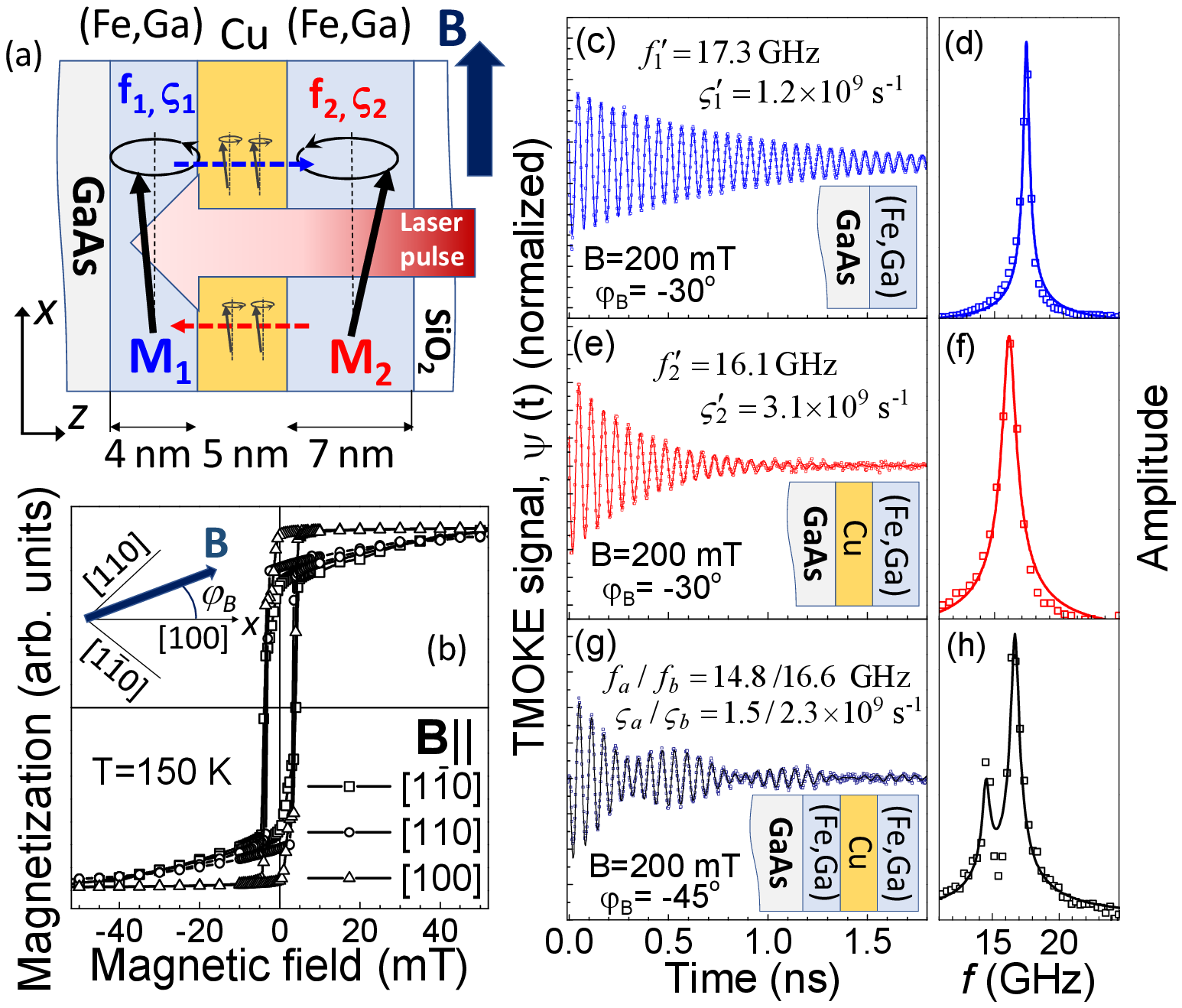}
 \caption{
 (a) Scheme of the studied PSV structure and the experimental idea. (b) SQUID magnetization curves measured for the three in-plane orientations of $\bm{\mbox{B}}$. Inset shows the used coordinate system. (c)-(f) TMOKE signals (left panels) and their FFT spectra (right panels) measured in a single Galfenol layer of 4-nm on GaAs (c,d) and a single 7-nm Galfenol layer on Cu (e,f). At the chosen azimuthal angle $\varphi_B=-30\degree$, MP with large amplitude is excited in both single layers. (g,h) TMOKE signal measured in PSV with no resonance of the precessing magnetizations. In (c)-(h), symbols show the experimental data; solid curves are fits with the parameters $f$ and $\zeta$ shown in the respective panels.
 }
\end{figure}

	Pump laser pulses (100-kHz repetition rate, 800-nm wavelength, 200-fs pulse duration, 10 mJ/cm$^2$ fluence excitation density within 100-$\mu$m focus spot) hit the PSV and launch MP by inducing ultrafast changes of the magnetic anisotropy \cite{38}. The laser penetration depth of 25~nm exceeds the total thickness of the PSV layer sequence. Thus, the pump excites both FM layers, thereby triggering simultaneously the precession of $\bm{\mbox{M}}_1$ and $\bm{\mbox{M}}_2$. The uncoupled precessions of $\bm{\mbox{M}}_1$ and $\bm{\mbox{M}}_2$ are characterized by the frequencies $f_{1,2}$ and decay rates $\zeta_{1,2}$.  Decay of MP occurs not only due to intrinsic processes, but also due to SP into the Cu-layer \cite{11,12}. The spin diffusion length in Cu exceeds significantly the spacer thickness \cite{40}, so that we expect the SC, pumped by the precessing magnetization in one layer, to exert an ac-torque on the magnetization of the other layer and thereby to affect its precession \cite{23,24,25,26}. Coupled modes should form close to the resonance $f_1 = f_2$ \cite {27,28}. To observe the coupling, we monitored the magnetization through the transient polar magneto-optical Kerr effect (TMOKE) in a pump-probe experiment. The rotation of the polarization plane, $\psi(t)$, of the linearly polarized probe beam focused to a spot of 60-$\mu$m diameter and reflected from the structure as a function of the time delay between the pump and probe pulses provides information about the temporal evolution of the total magnetization, $\bm{\mbox{M}}_1+\bm{\mbox{M}}_2$. Varying the external magnetic field we tuned the MP parameters:  $f_{1,2}$ and $\zeta_{1,2}$, as well as the contribution of SP to the magnetization dynamics.

      For comparison, we performed corresponding measurements on single Galfenol layers identical to those in the PSV. Figures ~1c and 1e show the $\psi (t)$ of these single layers
revealing 
exponentially decaying oscillations. Their fast Fourier transforms (FFTs) in Figs.~1d and 1f show single spectral lines with the MP parameters listed in each panel (hereafter primes indicate the single layer parameters). The much faster MP decay in the layer on top of Cu could be, for instance, due to 
SP into the Cu layer. The difference between $f^\prime_1$ and $f^\prime_2$ is due to different magnetic anisotropies: a weak cubic one in (Fe,Ga)/Cu and a stronger cubic anisotropy with additional uniaxial and out-of-plane contributions in (Fe,Ga)/GaAs \cite{37}.

      In the PSV both layers contribute to the measured MP. The corresponding TMOKE in Fig.~1g contains two oscillating components with different frequencies, as seen from the FFT spectrum (Fig.~1h).
The signal can be well described as a sum of two damped sine functions with two parameter sets indexed $a$ and $b$:
\begin{eqnarray}
\psi (t)= A_a \sin (2\pi f_a t-\phi_a) \exp (-\zeta_a t) + \nonumber\\
A_b \sin (2\pi f_b t-\phi_b) \exp (-\zeta_b t).
\end{eqnarray}
The fit to $\psi (t)$  in Fig. 1g yields  $f_a=14.8~$GHz, $\zeta_a = 1.5 \times 10^9~\mbox{s}^{-1}$, $f_b=16.6~$GHz, $\zeta_b = 2.3 \times 10^9~\mbox{s}^{-1}$. The solid line in Fig.~1h shows the FFT spectrum 
corresponding to the fit. Because the frequency splitting of the spectral peaks is larger than their widths, we attribute the two components to $\bm{\mbox{M}}_1$ and $\bm{\mbox{M}}_2$, both precessing at their individual frequencies, so that we may assign $f_{a,b}=f_{1,2}$ and $\zeta_{a,b} = \zeta_{1,2}$.
\begin{figure}
 \includegraphics[scale=0.84]{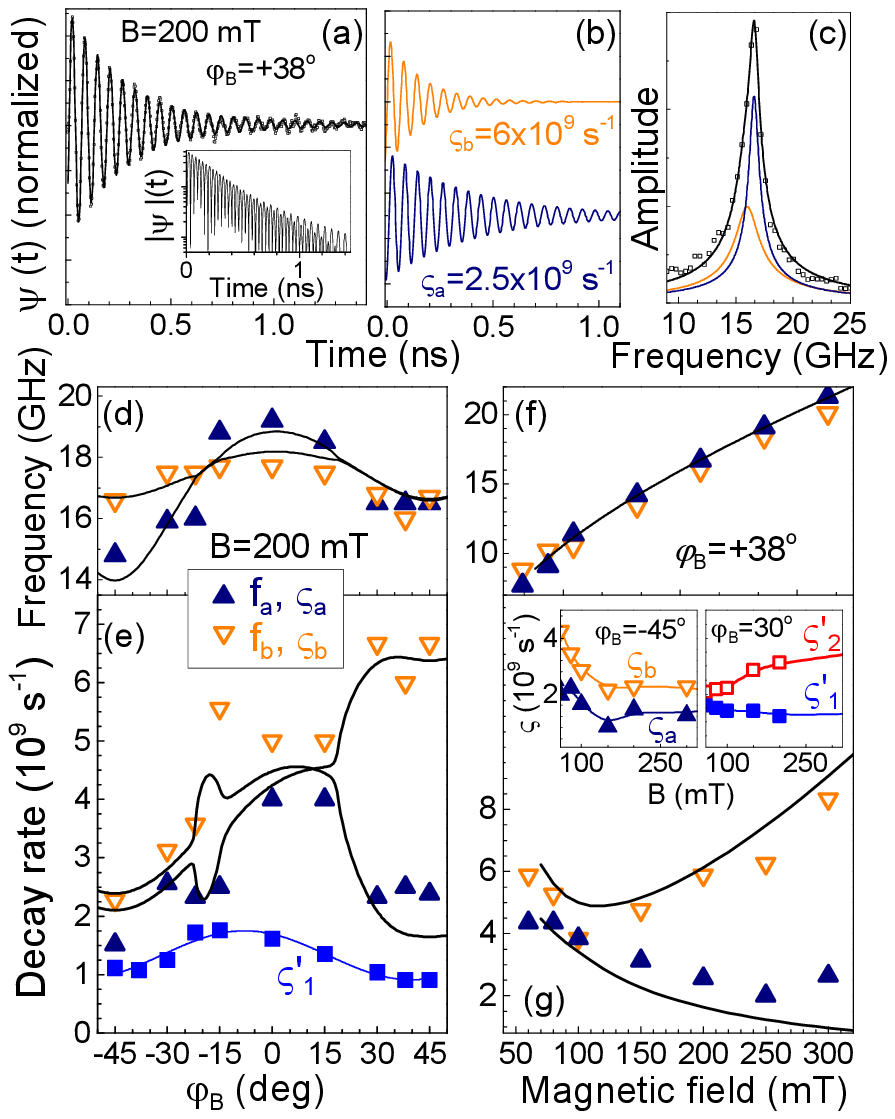}
 \caption{
(a) Experimental signal (symbols) measured in the PSV at resonant conditions and fit by Eq. (1) (solid line). The inset shows $|\psi (t)|$ in logarithmic scale. (b) Long- and short-living precessional modes contributing to $\psi (t)$ with respective decay rates, obtained by fitting by Eq. (1). (c) FFT spectra of the experimental signal (symbols), the fit (black solid line) and the long- and short-living modes (dark blue and orange lines, respectively). (d),(e) Azimuthal dependences of the precession frequencies $f_{a,b} (\varphi_B)$ [panel (d)] and the decay rates $\zeta_{a,b} (\varphi_B)$  [panel (e)] at $B=200$~mT. (f), (g) Field dependences $f_{a,b}(B)$ [panel (f)] and $\zeta_{a,b} )(B)$ [panel (g)] at $\varphi_B =+38\degree$. The insets in panel (g) show the field dependences of the decay rates $\zeta_{a,b}(B)$ measured in the PSV at $\varphi_B=-45\degree$ (left inset) and $\zeta^\prime_{1,2} (B)$ in the single layer structures at $\varphi_B=+30\degree$ (right inset). In panels (d)-(g) the values obtained from the experiment are shown by symbols; the lines show the calculated dependences. In the insets the lines are guides for the eye.
 }
\end{figure}

      Owing to the different magnetic anisotropies of Layers~1 and 2 we can change the detuning of the precession frequencies by varying the angle $\varphi_B$. Figure~2a shows $\psi(t)$ measured at $B=200~$mT applied at $\varphi_B=+38\degree$. Contrary to the case where $\varphi_B=-45\degree$, here we can neither separate the signal into two independent oscillations with different frequencies, nor describe it as single-frequency oscillation with mono-exponential decay. The inset of Fig. ~2a showing the absolute $\psi (t)$ on a logarithmic scale clearly indicates two decay rates of $\psi (t)$. The analysis shows that $\psi (t)$ is the sum of two components (see Fig. 2b) with close frequencies, $f_a\approx f_b \approx 16$~GHz, but significantly different decay rates: $\zeta_a \approx 2.5\times 10^9~\mbox{s}^{-1}$  and $\zeta_b \approx 6\times 10^9~\mbox{s}^{-1}$. The FFT spectrum in Fig. 2c is fitted well by two spectral lines centered at $f\approx 16~$GHz, one narrow and one broad.  This result is our main experimental observation. Further analysis of the field dependences of $f_{a,b}$ and $\zeta_{a,b}$ proves that this effect is due to collective precession of $\bm{\mbox{M}}_1$ and $\bm{\mbox{M}}_2$, coupled by SP.

      Figure~2d shows
$f_{a,b}(\varphi_B)$ at $B=200~$mT, from which we identify two dependences corresponding to the expected magnetic anisotropies:  $f_a(\varphi_B)$ complies with a cubic anisotropy plus an uniaxial distortion as observed in the single 4-nm Fe$_{0.81}$Ga$_{0.19}$ layer on GaAs; $f_a(\varphi_B)$ agrees with the weak cubic anisotropy of the 7-nm Fe$_{0.81}$Ga$_{0.19}$ layer on Cu. At any tested direction of $\bm{\mbox{B}}$, the best fit of the data gives two frequencies contributing to the TMOKE signal, though at some angles (e.g. $\varphi_B > 30\degree$) they have very close values.

      Contrary to the precession frequencies, the decay rates in Fig.~2e do not demonstrate a behavior corresponding to precessions in single layers.
For $\varphi_B \approx -15 \degree$ and $+30\degree <\varphi_B < +45\degree$, where the precession frequencies almost coincide, we observe a pronounced splitting of the decay rates as shown above for $\varphi_B = +38\degree$. We obtain $\zeta_a \approx 2.5\times 10^9~ \mbox{s}^{-1}$  and $\zeta_b \approx 5.5\times 10^9~ \mbox{s}^{-1}$  at $\varphi_B \approx -15\degree$, and  $\zeta_a \approx 2\times 10^9~ \mbox{s}^{-1}$  and $\zeta_b \approx 7\times 10^9~ \mbox{s}^{-1}$  for $+30\degree < \varphi_B < +45\degree$. For comparison, $\zeta^\prime_1 (\varphi_B)$  measured on the Fe$_{0.81}$Ga$_{0.19}$/GaAs structure (see the blue symbols in Fig. ~2e) shows a smooth variation around a broad maximum at $\varphi_B=0\degree$, without the abrupt changes observed for the PSV.

      We examine also the field dependences of the precession parameters at a fixed direction of $\bm{\mbox{B}}$, where the resonance condition is fulfilled. Figs.~2f and 2g show $f_{a,b} (B)$ and $\zeta_{a,b}(B)$  at $\varphi_B=+38\degree$. Across the scanned range of $B$ the two components contributing to $\psi (t)$ have closely matched frequencies, while their decay rates show a pronounced increasing splitting at $B>100$~mT.
This behavior is different to the dependences of $\zeta_{a,b} (B)$  at $\varphi_B=-45\degree$  where no resonance is present and we find a small, field-independent difference between the two decay rates, see the left inset of Fig.~2g. The dependence $\zeta_{a} (B)$ at $\varphi_B=+38\degree$  agrees with the dependences of $\zeta^\prime _{1}(B)$ in the single layer shown in the right inset of Fig.~2g (though, measured at a slightly smaller $\varphi_B$  when both layers exhibit large precession amplitudes). However, $\zeta_b$ increases with $B$ much faster than  $\zeta^\prime _2$. This is an indication of the SP contribution to the dacay of this mode.

      The dependences of the decay rates on magnetic field as well as their splitting at resonance are well explained by dynamic coupling of the two magnetizations by SP. Figure~3a shows the suggested coupling mechanism. The precessions of $\bm{\mbox{M}}_1$  and $\bm{\mbox{M}}_2$ decay through two main channels: intrinsic damping characterized by the coefficients   $\alpha_1$ and $\alpha_2$, and SP into the Cu spacer characterized by  $\beta_1$ and $\beta_2$. The pure SC generated by the precessing $\bm{\mbox{M}}_1$ exerts an ac-spin torque on $\bm{\mbox{M}}_2$ affecting its precession, and vice versa. The temporal evolutions of $\bm{\mbox{M}}_1$ and $\bm{\mbox{M}}_2$ coupled by this dissipative mechanism can be described by the modified Landau-Lifshitz-Gilbert equations \cite{27}:
\begin{eqnarray}
\frac{d \bm{\mbox{M}}_{i}}{dt}= \gamma_g \bm{\mbox{B}}_{eff}^{(i)} \times   \bm{\mbox{M}}_{i} + \\  \frac{\alpha_i+\beta_i}{M} \bm{\mbox{M}}_{i}\times \frac{d \bm{\mbox{M}}_{i}}{dt} - \nonumber
\frac{\beta_j}{M} \bm{\mbox{M}}_{j}\times \frac{d \bm{\mbox{M}}_{j}}{dt},
\end{eqnarray}
where $\gamma_g$ is the gyromagnetic ratio, $\bm{\mbox{B}}_{eff}^{(i)}$ is the effective magnetic field determined by the magnetic anisotropy and the applied magnetic field, $i$ and $j$ denote the magnetic layers ($i\neq j$). Solution of the linearized version of Eq. (2) yields coupled precessional modes. If the precession frequencies of the individual layers are well separated and the coupling is weak $\zeta_{1,2} \approx \alpha_{1,2} + \beta_{1,2}$. However, close to resonance, when the frequency splitting is smaller than the average widths of the precessional modes, the magnetizations precess with the same frequencies, but show a double-exponential decay, representing a superposition of two modes with decay rates:
$$
\zeta_{a,b} \sim  \frac{\alpha_1\!+\!\beta_1\!+\!\alpha_2\!+\!\beta_2 \mp   \sqrt{4 \beta_1 \beta_2\!+\!(\alpha_1\!+\!\beta_1\!-\!\alpha_2\!-\!\beta_2)^2}}{2}.
$$
The difference between the damping parameters for the two coupled modes due to SP is illustrated in Fig. 3a. The long-living mode with suppressed damping can be considered as the two magnetizations precessing in-phase. Then the spin torques from the two magnetizations support the joint precession. The damping of this mode, $\zeta_a$, is close to the intrinsic one. The short-living mode, in contrast, represents counter-phase precession of $\bm{\mbox{M}}_1$ and $\bm{\mbox{M}}_2$, which causes a mutual damping. Approximately, $\zeta_a-\zeta_b \sim 2\sqrt{\beta_1\beta_2}$.

 \begin{figure}
 \includegraphics[scale=0.66]{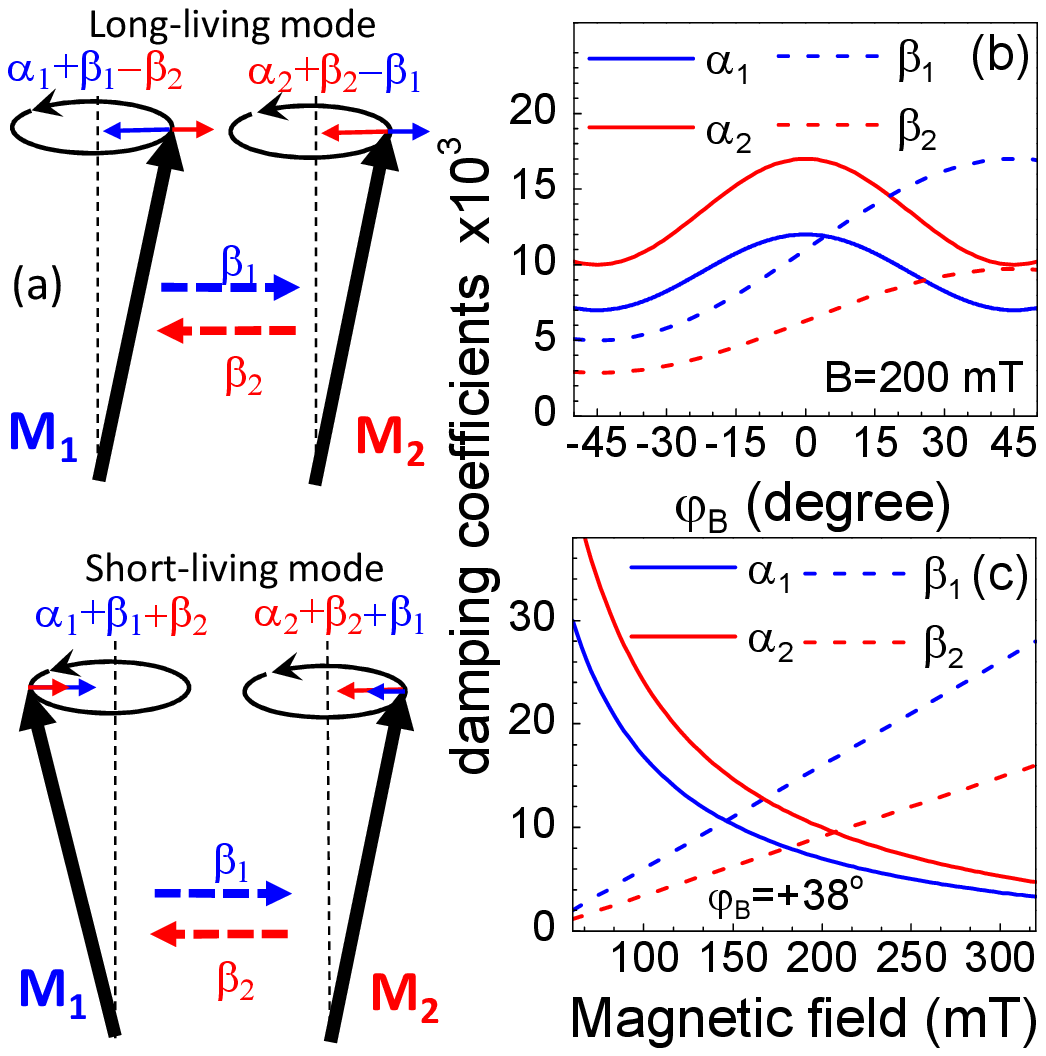}
 \caption{
(a) Qualitative and simplified description of the two collective precessional modes mediated by the SP. Upper sketch demonstrates the long-living mode, in which the reciprocal spin currents generate spin torques supporting the precession. Lower panel shows the short-living mode, in which the contribution of the spin currents is destructive for the precession. (b,c) Dependences of the damping coefficients $\alpha_{1,2}$ and $\beta_{1,2}$ on $\varphi_B$ (panel b) and $B$ (panel c) obtained from modeling of the experimental dependences.
 }
\end{figure}

      To substantiate our interpretation, we modeled the magnetization kinetics in the PSV numerically. The solid curves in Figs.~2d-2g give the calculated results using the following magnetic anisotropy parameters: $K_1^{(1)} = 20$~mT, $K_\perp^{(1)} = -40$~mT, $K_{||}^{(1)} = 20$~mT for the bottom 4-nm Galfenol layer, and  $K_1^{(2)} = 8$~mT, $K_\perp^{(2)} = -65$~mT, $K_{||}^{(2)} = 0$~mT, for the top one. The parameters $K_{1,\perp,||}$ represent the cubic, perpendicular and in-plane uniaxial anisotropies, respectively. The magnetization was taken to be $\mu_0 M =1.59$~T \cite{41}. The angular and field dependences of the coefficients $\alpha_{1,2}$ and $\beta_{1,2}$  providing the best agreement with the experimental data are summarized in Figs. 3b and 3c. The dependences of $\alpha_{1,2}$ correspond well to the  $\zeta^\prime_{1,2} (\varphi_B,B)$ for the single layers, though in the PSV the absolute values are a bit larger. The dependence of $\beta_{1,2}$ in Fig. 3b demonstrates a pronounced angular anisotropy of the SP efficiency. Indeed, in the PSV the decay at $\varphi_B =-45\degree$ for both layers is about the same as that of the long-living mode at $\varphi_B >25\degree$, which is close to the intrinsic one. Since in the single 4-nm Galfenol layer the decay rates for $\varphi_B =\pm 45\degree$ are similar, the SP contribution to the decay for $\varphi_B =-45\degree$ is marginal. Note that a strong SP anisotropy in a PSV with in-plane magnetic anisotropy of one layer had been reported in \cite{25}.
The SP constants also depend on magnetic field as seen in Fig.~3c. Indeed, for the selected $\varphi_B =+38\degree$, the precession frequencies are close to resonance across the whole range of magnetic fields, and we always observe coupled modes. Thus, the observed increase of the decay rate splitting with $B$ suggests a corresponding dependence of $\beta_{1,2} (B)$. This agrees with the dependence of $\zeta^\prime_2 (B)$ measured in the Fe$_{0.81}$Ga$_{0.19}$/Cu/GaAs structure and shown in the right inset of Fig.~2g. The twice larger splitting of $\zeta_a$ and $\zeta_b$ (Fig. 2g, main panel) confirms the formation of a collective precessional mode.

      It is interesting to note that the demonstrated collective precession of two magnetizations mediated by SP is a rare example of pure dissipative coupling, which in a quantum-mechanical approach would be described by a non-Hermitian matrix \cite{42}. This coupling regime for two oscillators results in the formation of two degenerate modes with split decay rates. Although realization of dissipative coupling promises interesting effects, in partucular in nano-optomechanical structures \cite{43},  the number of systems with such a coupling is limited so far \cite{44,45}. To the best of our knowledge, precessing magnetizations dynamically coupled by SP have not been considered in this context. Indeed, in FMR experiments on similar structures, the magnetizations are driven by microwaves, which precisely set the precession phase for both magnetizations. This results in the observation of only one collective mode: either the long-living mode for parallel magnetizations \cite{27,28} or the short-living mode if the magnetizations are antiparallel \cite{28}. The pulsed optical excitation in our experiment triggers instantly the precession of the two magnetizations, and the initial precession phases are determined by the anisotropy parameters of the layers. This allows us to observe both modes in the collective magnetization dynamics.

      To conclude, we demonstrated that ultrafast optical excitation of the magnetization precession is a powerful tool for triggering pure spin currents in ferromagnetic multilayer structures without the need for applying microwaves. For our pseudo spin-valve this was confirmed by the observation of collective precessional modes dissipatively coupled by the spin pumping. The optical excitation allows one to launch a superposition of these modes over a wide frequency range not achievable for microwave driving. The use of Galfenol-based spin valves allows also designing of a complex spin current temporal pattern by resonant phonon driving of the magnetization precession in a spin-valve structure inserted into a phononic nanoresonator \cite{46}.


We are grateful to Andrey Akimov, Davide Bossini, and  Andrew Armour for fruitful discussions. This work was supported by the Deutsche Forschungsgemeinschaft in the frame of the International Collaborative Research Center TRR160 [project B6], and by the Engineering and Physical Science Research Council [grant no. EP/H003487/1] through support for the growth and characterization of the Galfenol-based nanostructures in the University of Nottingham. The experimental studies in the Ioffe Institute were performed under support of the Russian Scientific Foundation [grant no. 16-12-10485]. The Volkswagen Foundation supported the cooperative work with Lashkarev Institute [grant no. 90418].

\end{document}